\documentclass[twocolumn,amssymb,aps,prd]{revtex4}
\usepackage{epsfig}

\newcommand{\astar}{\ensuremath{a_*}}

\newcommand{\omtil}{\ensuremath{\widetilde{\omega}}}
\newcommand{\Qtil}{\ensuremath{\widetilde{Q}}}

\newcommand{\rH}{\ensuremath{r_h}}

\newcommand{\bmax}{\ensuremath{b_\mathrm{max}}}

\begin{document}

\hspace{8cm}\hbox to\hsize{\hfill KIAS-P03086}
 \vspace{0.9cm}
\hspace{9cm}\hbox to \hsize{\hfill TUM-HEP-535-03}

\title{Black Holes at future colliders
\footnote{This paper is based on talks by SC in several
conferences. They include DESY theory workshop in DESY, Hamburg,
Germany Sep.23-26, 2003, International conference on Gravitation
and Astrophysics of APCTP(ICGA-VI) in Ewha Woman's University,
Seoul, Korea, Oct. 6-9, 2003 and International conference on
Flavour physics (ICFP-II) in KIAS, Seoul, Korea, Oct. 6-11, 2003.}
}%

\author{Daisuke Ida}%
\email{d.ida@th.phys.titech.ac.jp} \affiliation{Department of
Physics, Tokyo Institute of Technology,  Tokyo 152-8551, Japan}
\author{Kin-ya Oda}%
\email{odakin@ph.tum.de} \affiliation{Physik Dept.\ T30e, TU
M\"unchen,  James Franck Str., D-85748 Garching, Germany}
\author{Seong Chan Park}%
\email{spark@kias.re.kr} \affiliation{Korea Institute for Advanced
Study (KIAS), Seoul 130-012, Korea}

\begin{abstract} We consider the production and
decay of TeV sized black holes. After discussing  evaluation of
the production cross section of higher dimensional rotating black
holes and black rings, the master equation for general spin-$s$
fields confined on brane world is derived. For five-dimensional
(Randall-Sundrum) black holes, we obtain analytic formulae for the
greybody factors in low frequency expansion.
\end{abstract}

\preprint{<hep-ph/0311297>}
\maketitle 

\section{Introduction}
The scattering process of two particles at CM energies in the
trans-Planck domain, is well calculable using known laws of
physics, because gravitational interaction dominates over all
other interactions. Non-trivial quantum gravitational (or string/M
theoretical) phenomena are well behind the horizon
\cite{'tHooft:1987rb}. If the impact parameter is less than the
black hole radius corresponding to the CM energy then one
naturally expects a black hole to form.
When nature realizes TeV scale gravity
scenario~\cite{Arkani-Hamed:1998rs,Antoniadis:1998ig,Randall:1999ee},
one of the most intriguing prediction would be copious production
of TeV sized black holes at near future particle colliders and in
ultra high energy cosmic rays
\cite{Banks:1999gd,Argyres:1998qn,Emparan:2000rs,Giddings:2001bu,Dimopoulos:2001hw}.
(See also some recent papers~\cite{BHcollider} for particle
accelerator signals and \cite{UHEneutrino,Anchordoqui:2001cg} for
cosmic ray signals.)
The production cross section of black hole in the higher
dimensional case was obtained in ref.~\cite{Ida:2002ez} under the
assumptions of ``Hoop conjecture''~\cite{Thorne:1972} by taking
angular momenta into account  and the result has been numerically
proved in refs.~\cite{Eardley:2002re,Yoshino:2002tx}(See Ref.
\cite{Park:2001xc} and also \cite{Anchordoqui:2001cg} where
similar analysis were made to estimate the cross-section by taking
angular momenta into account.).
Once produced, black holes lose its masses and angular momenta
through the Hawking radiation \cite{Hawking:1975sw}. The Hawking
radiation is determined for each mode by the greybody factor,
i.e.\ the absorption probability  of an in\-com\-ing wave of the
corresponding mode.  For four dimensional case, it is first
calculated for spin~0 field by A. A.
Starobinsky~\cite{Starobinsky:I}, then for spin~1, ~2 and ~1/2
fields by S. A. Teukolsky  and Don
N.Page~\cite{Teukolsky:1972my,Starobinsky:II,Teukolsky:I,Teukolsky:II,Teukolsky:III,Page:1976df,Page:1976ki}.
The absorption cross section of a non-rotating BH for all
frequencies and with an analytic expression was computed by N.
Sanchez \cite{Sanchez:1978si}.

The master equation for general brane-fields with arbitrary
spin-$s$ was obtained in ref.~\cite{Ida:2002ez} for rotating black
holes in higher dimensional spacetime and its non-rotating limit
was confirmed in ref.~\cite{Harris:2003eg}.
Analytic expressions of greybody factors for rotating black holes
were obtained in five dimensional (Randall-Sundrum) case in
ref.~\cite{Ida:2002ez} and also for non-rotating limit in the
series of papers ~\cite{Kanti:2002ge,Kanti:2002nr}.

\section{Production of rotating black holes}
First we briefly review the properties of the rotating $(4+n)$-
dimensional black hole~\cite{Myers:1986un}. In general, higher
dimensional black hole may have $\lfloor(n+3)/2\rfloor$ angular
momenta. When the black hole is produced in the collision of two
particles on the brane, where the initial state has only single
angular momentum, it is sufficient to consider that the only
single angular momentum is non-zero.

\begin{figure}
\begin{center}
\leavevmode \epsfig{file=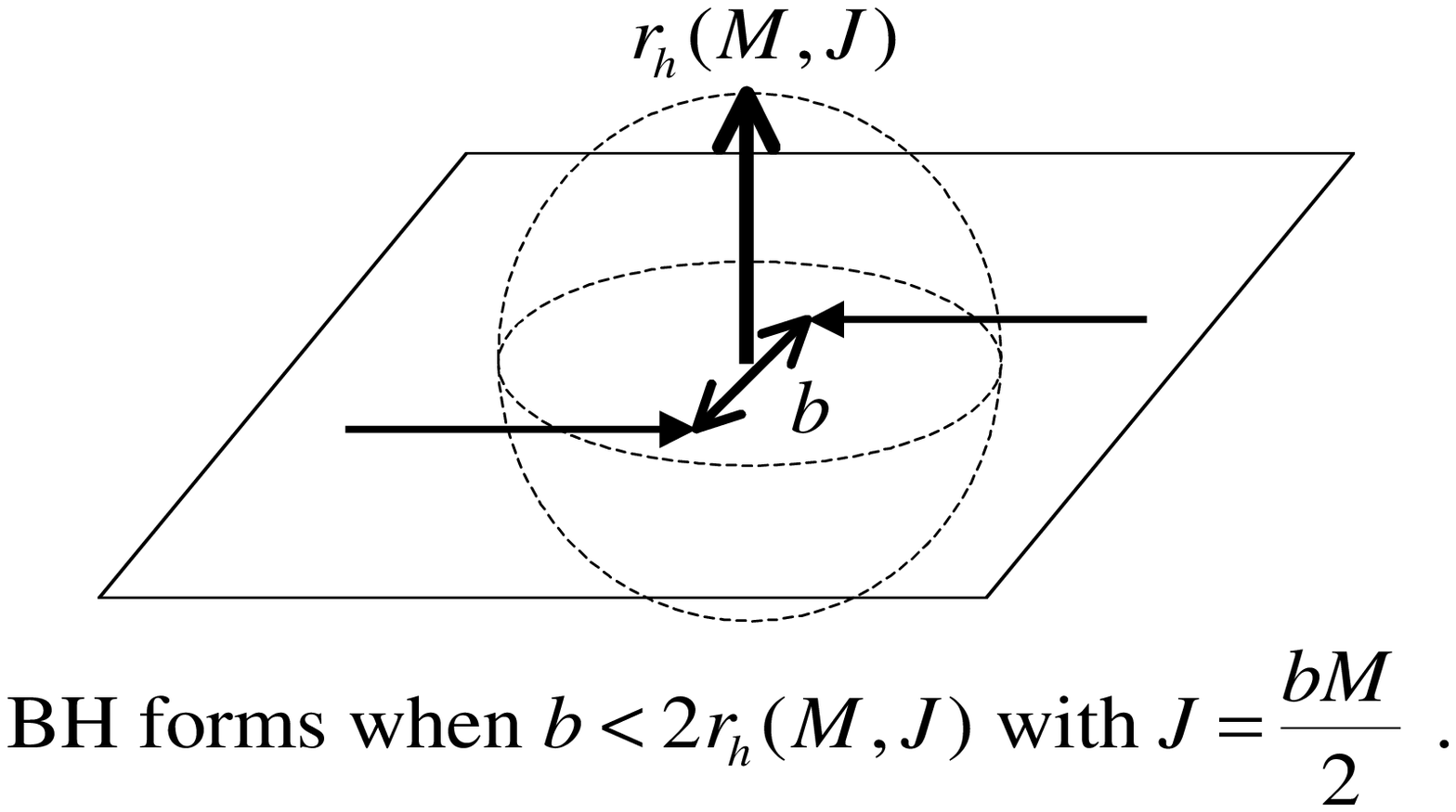,width=0.5\textwidth
}\vspace{-2.4cm} \caption{ Schematic picture for the condition of
the black hole formation. } \label{fig:BHpicture}
\end{center}
\end{figure}
 In the
Boyer-Lindquist coordinate, the metric for the black hole with
single angular momentum takes the following
form~\cite{Myers:1986un}
\begin{eqnarray}
g &=&\left(1-\frac{\mu r^{-n+1}}{\Sigma(r,\vartheta)}\right)dt^2
\nonumber
\\
  &&-\sin^2\vartheta\left(
    r^2+a^2+a^2\sin^2\vartheta\frac{\mu r^{-n+1}}{\Sigma(r,\vartheta)}\right)d\varphi^2
    \nonumber\\
&&\mbox{}+2a\sin^2\vartheta\frac{\mu
r^{-n+1}}{\Sigma(r,\vartheta)}dtd\varphi
  -\frac{\Sigma(r,\vartheta)}{\Delta(r)}dr^2-\Sigma(r,\vartheta) d\vartheta^2\nonumber\\
&&\mbox{}-r^2\cos^2\vartheta\,d\Omega^n, \label{eq:metric_general}
\end{eqnarray}
where
\begin{eqnarray}
\Sigma(r,\vartheta)&=&r^2+a^2\cos^2\vartheta,\nonumber\\
\Delta(r)       &=&r^2+a^2-\mu r^{-n+1}.\nonumber
\end{eqnarray}
We can see that the horizon occurs when $\Delta(r)=0$, i.e.\ when
$r=r_h$ with
\begin{eqnarray}
r_h &=& \left[\frac{\mu}{1+\astar^2}\right]^{1/(n+1)} =
(1+\astar^2)^{-1/(n+1)}r_S,\label{eq:r_h}
\end{eqnarray}
where $\astar=a/r_h$ and $r_S$ is the Schwarzschild radius for
given mass. Note that there is only single horizon when $n\geq 1$
(contrary to the four-dimensional Kerr black hole which has inner
and outer horizons) and its radius is independent of the angular
coordinates.
We can obtain the 
total mass
 $M$ and angular momentum $J$ of the black hole
from the metric (\ref{eq:metric_general})
\begin{eqnarray}
M = \frac{(n+2)A_{n+2}}{16\pi G}\mu, \hspace{1cm} J =
\frac{2}{n+2}Ma,\label{eq:MandJ}
\end{eqnarray}
where $A_{n+2}$$=$$2\,\pi^{(n+3)/2}/\Gamma(\frac{n+3}{2})$ is the
area of unit sphere $S^{n+2}$ and $G$ is  the $(4+n)$-dimensional
Newton constant. Therefore we may consider $\mu$ and $a$ (or
$r_h^{-1}$ and $\astar$) as the normalized mass and angular
momentum parameters, respectively. We note that there are no upper
bound on $a$ when $n\geq 2$ nor on $\astar$ when $n\geq 1$,
contrary to the four-dimensional case where both $a$ and $\astar$
are bounded from above. We concentrate on the brane field
equations and hence only the induced metric on the brane is
relevant, where the last term in eq.~(\ref{eq:metric_general})
vanishes and the angular variables $\vartheta$ and $\varphi$ are
redefined to take the values $0\leq\vartheta\leq\pi$ and
$0\leq\varphi<2\pi$.

\subsection{Production cross section}
We estimate the production cross section of rotating black holes
within the classical picture.
Let us consider a collision of two massless particles with finite
impact parameter $b$ and CM energy $\sqrt{s}=M_i$ so that each
particle has energy $M_i/2$ in the CM frame.(see
Fig.~\ref{fig:BHpicture} for schematic picture)
The initial angular momentum before collision is $J_i=bM_i/2$ in
the CM frame. Suppose that a black hole forms whenever the initial
two particles can be wrapped inside the event horizon of the black
hole with the mass $M=M_i$ and angular momentum $J=J_i$ , i.e.,
when
\begin{eqnarray}
b &<& 2r_h(M,J)=2r_h(M_i,bM_i/2), \label{eq:our_condition}
\end{eqnarray}
where $r_h(M,J)$ is defined through eqs.~(\ref{eq:r_h}) and
(\ref{eq:MandJ}). Since the right hand side is monotonically
decreasing function of $b$, there is maximum value $\bmax$ which
saturates the inequality~(\ref{eq:our_condition})
\begin{eqnarray}
\bmax(M)=2\left[1+\left(\frac{n+2}{2}\right)^2\right]^{-{1\over
n+1}}r_S(M), \label{eq:bmax_formula}
\end{eqnarray}
where $r_S(M)$ is defined by $r_S(M)=\mu(M)^{1/(n+1)}$ and
eq.~(\ref{eq:MandJ}). When $b=\bmax$, the rotation parameter
$\astar$ takes the maximal value $(\astar)_\mathrm{max}=(n+2)/2$.

The formula (\ref{eq:bmax_formula}) fits the numerical result of
$\bmax$ with full consideration of the general relativity
by 
Yoshino and Nambu~\cite{Yoshino:2002tx} within the accuracy less
than 1.5\% for $n\geq 2$ and 6.5\% for $n=1$:
\begin{eqnarray}
\begin{array}{c|cccccccc}
n  & 1 & 2 & 3 & 4 & 5 & 6 & 7 \\
\hline
R_\mathrm{Numerical}~
   & 1.04 & 1.16 & 1.23 & 1.28 & 1.32 & 1.35 & 1.37\\
R_\mathrm{Analytic}
    & 1.11 & 1.17 & 1.22 & 1.26 & 1.30 & 1.33 & 1.36
\end{array} \nonumber
\end{eqnarray}
where $R$ denotes $R=\bmax/r_S(M)$.

Our result is obtained in the approximation that we neglect all
the effects by the junk emissions in the balding phase and hence
that the initial CM energy $M_i$ and angular momentum $J_i$ become
directly the resultant black hole mass $M=M_i$ and angular
momentum $J=J_i$. The coincidence of our result with the numerical
study~\cite{Yoshino:2002tx}
suggests that 
this approximation would be actually viable for higher dimensional
black hole formation at least unless $b$ is very close to $\bmax$.

Once we neglect the balding phase, the initial impact parameter
$b$ directly leads to the resultant angular momentum of the black
hole $J=bM/2$. Since the impact parameter $[b, b+db]$ contributes
to the cross section $2\pi bdb$, this relation between $b$ and $J$
tells us the (differential) production cross section of the black
hole with its mass $M$ and its angular momentum in $[J,J+dJ]$
\begin{eqnarray}
d\sigma(M,J)=\left\{
   \begin{array}{cc}
   8\pi JdJ/M^2 & (J<J_\mathrm{max})\\
   0            & (J>J_\mathrm{max})
   \end{array}\right.
, \label{eq:dsigma_dJ}
\end{eqnarray}
where
\begin{eqnarray}
J_\mathrm{max}
   &=&\frac{\bmax M}{2}=j_n\,\left(\frac{M}{M_P}\right)^{n+2\over n+1}
\end{eqnarray}
with
\begin{eqnarray}
j_n&=&\left[2^n\pi^{n-3\over 2}\Gamma\left(n+3\over 2\right) \over
(n+2)
   \left[1+\left(n+2\over
   2\right)^2\right]\right]^{1/(n+1)},\nonumber \\
 M_P&=&\left((2\pi)^n\over 8\pi G\right)^{1/(n+2)}.
\end{eqnarray}

It is observed that the differential cross
section~(\ref{eq:dsigma_dJ}) linearly increases with the angular
momentum.
We expect that this behavior is correct as the first
approximation, so that
the black holes tend to be produced with larger angular momenta. 
At the typical LHC energy $M/M_P=5$, the value of $J_\mathrm{max}$
is 
$J_\mathrm{max}=2.9, 4.5, \ldots, 10, 12$\ for $n=1, 2, \ldots,
6,7$, respectively. This means that the semi-classical treatment
of the angular momentum becomes increasingly valid for large $n$.

Integrating the expression~(\ref{eq:dsigma_dJ}) simply gives
\begin{eqnarray}
\sigma(M) &=&\pi\bmax^2 \nonumber \\
   &=&4\left[1+\left(\frac{n+2}{2}\right)^2\right]^{-2/(n+1)}\,\pi r_S(M)^2
   \nonumber\\
&=&F\,\pi r_S(M)^2.
\end{eqnarray}
The form factor $F$ \footnote{In Ref.~\cite{Anchordoqui:2001cg},
$\bmax$ was assumed to be given by $\bmax=r_h(M,M\bmax/2)$ rather
than Eq.~(\ref{eq:our_condition}), resulting in the form factor
less than 1: $F\simeq 0.62$--0.64 in any dimensions $1\leq n\leq
7$.}  is summarized as
\begin{eqnarray}
\begin{array}{c|cccccccc}
n  & 1 & 2 & 3 & 4 & 5 & 6 & 7 \\
\hline F_\mathrm{NY}~
   & 1.084 & 1.341 & 1.515 & 1.642 & 1.741 & 1.819 & 1.883\\
F_\mathrm{Our}
   & 1.231 & 1.368 & 1.486 & 1.592 & 1.690 & 1.780 & 1.863
\end{array}. \label{eq:Rsq_table} \nonumber
\end{eqnarray}

%
%
%
This result implies that  we would underestimate the production
cross section of black holes if we did not take the angular
momentum into account and that it becomes more significant for
higher dimensions. We point out that this effect has been often
overlooked in the literature.


\subsection{Rotating black ring}
A higher-dimensional black hole can have various nontrivial
topology, and the uniqueness property of stationary black holes
fails in five (and probably in higher) dimensions. The typical
example in five dimensions has been recently given by Emparan and
Reall~\cite{Emparan:2001wn}. They have explicitly provided a
solution of the five-dimensional vacuum Einstein equation, which
represents the stationary rotating black ring (homeomorphic to
$S^1\times S^2$). In this case, the centrifugal force prevents the
black ring from collapsing. When the angular momentum is not large
enough, the black ring will collapse to the Kerr black hole due to
the gravitational attraction and some effective tension of the
ring source. In fact, this five dimensional black ring solution
has the minimum possible value of the angular momentum given by
\begin{eqnarray}
J_\mathrm{min}=k_\mathrm{BR}\left(M\over M_P\right)^{3/2},
\end{eqnarray}
where $k_\mathrm{BR}=0.282$. On the other hand, we have the upper
bound for the angular momentum of the black holes produced by
particle collisions:
\begin{eqnarray}
J_\mathrm{max}=j_1\left(M\over M_P\right)^{3/2},
\end{eqnarray}
where $j_1=0.256$. Since these numerical values are of the same
order,
we cannot conclude the possibility of black ring productions at
colliders.

Now we consider the possibility of the higher dimensional black
rings, which is homeomorphic to, say, $S^1\times S^{n+1}$.
Corresponding Newtonian situation will be the system of a rotating
massive circle.  For simplicity, we just consider the
gravitational attraction and the centrifugal force of the massive
circle and neglect the effect of tension. Let $\ell$, $M$ and $J$
be the radius, the mass and the angular momentum
 of the massive circle.
The the minimum value of the angular momentum for exploding black
ring is estimated:
\begin{eqnarray}
J &\gtrsim& J_\mathrm{min} =k_n\left(M\over
M_P\right)^{(n+2)/(n+1)},
\end{eqnarray}
where
\begin{eqnarray}
k_n &=& 2^{-{2n^2+3n+7\over 2(n+1)}}\pi^{(n+6)(n-1)\over 4(n+1)}
\left[\Gamma\left(n+2\over 2\right)\over n+1\right]^{-{n-1\over
2(n+1)}}.
\end{eqnarray}

 $J_\mathrm{min}$
for exploding black rings is one or two order(s) of magnitude
smaller than $J_\mathrm{max}$ for collision limit when $n$ is
large.
Therefore we expect that the exploding black rings are possibly
produced at colliders if there are many extra dimensions, though
they will suffer from the black string instability when they
become sufficiently large thin rings.

\section{Radiations from rotating black hole}
In this section, we study the Hawking
radiation~\cite{Hawking:1975sw} from the higher dimensional Kerr
black hole~\cite{Myers:1986un}. The Hawking radiation is thermal
but not strictly black body due to the frequency dependent
greybody factor $\Gamma$, which is identical to the absorption
probability (by the hole) of the corresponding
mode~\cite{Hawking:1975sw,Page:1976df}. The quantity $1-\Gamma$
for each mode can be computed from the solution (to the wave
equation of that mode) which has no outgoing flux at the horizon
as the ratio of the incoming and outgoing flux at infinity.

\subsection{Brane field equations}
We derive the wave equations of the brane modes using the induced
four dimensional metric of the $(4+n)$-dimensional rotating black
hole~\cite{Myers:1986un}. The wave equations can be understood as
generalization of the Teukolsky
equation~\cite{Teukolsky:1972my,Teukolsky:I,Teukolsky:II,Teukolsky:III}
to the higher dimensional Kerr geometry. The derivation is shown
in Appendix.

We present the brane field equations for massless spin $s$ field
which are obtained from the metric (\ref{eq:metric_general}) with
the standard decomposition
\begin{eqnarray}
\Phi_s=R_s(r)S(\vartheta)e^{-i\omega t+im\varphi},
\end{eqnarray}
utilizing the Newman-Penrose formalism~\cite{Newman:1962qr}
\begin{eqnarray}
&&{1\over\sin\vartheta}{d\over d\vartheta}\left(\sin\vartheta {d
S\over d\vartheta}\right) +[ (s-a\omega\cos\vartheta)^2 \nonumber \\
&&-(s\cot\vartheta+m\csc\vartheta)^2 -s(s-1)+A
]S=0,\nonumber\\
\label{eq:angular}\\
%
%
&&\Delta^{-s}{d\over dr}\left(\Delta^{s+1}{dR\over dr}\right)
\nonumber \\
&&{}+\Biggl[ {K^2\over\Delta} +s\left( 4i\omega r -i{
\Delta_{,rr}K\over\Delta} +\Delta,_{rr}-2\right)
\nonumber\\
&&{} +2ma\omega-a^2\omega^2-A \Biggl]R=0, \label{eq:Teukolsky}
\end{eqnarray}
where
\begin{eqnarray}
K          &=&(r^2+a^2)\omega-ma.
\end{eqnarray}

The solution of eq.~(\ref{eq:angular}) is called spin-weighted
spheroidal harmonics ${}_sS_{lm}$ (see e.g.\
ref.~\cite{Teukolsky:II,Fackerell:1977}).

\subsection{Hawking radiation and greybody factor}
Since we have shown that massless brane field equations are
separable into radial and angular parts, we may write down the
power spectrum of the Hawking radiation~\cite{Hawking:1975sw} for
each massless brane mode
\begin{eqnarray}
\frac{dE_{s,l,m}}{dt\,d\omega\,d\varphi\,d\cos\vartheta}&=&
   \frac{\omega}{2\pi}
   \frac{{}_s\Gamma_{l,m}}{e^{\frac{\omega-m\Omega}{T}}\mp 1}
   \left|{}_sS_{lm}\right|^2,
   \label{eq:power_spectrum}
\end{eqnarray}
where $T$ and $\Omega$ are the Hawking temperature and the angular
velocity at the horizon, respectively given by
\begin{eqnarray}
T=\frac{(n+1)+(n-1)\astar^2}{4\pi(1+\astar^2)r_h}, \hspace{1cm}
\Omega=\frac{\astar}{(1+\astar^2)r_h},
\end{eqnarray}
and ${}_s\Gamma_{l,m}(r_h,a;\omega)$ is the greybody
factor~\cite{Hawking:1975sw,Page:1976df} which is identical to the
absorption probability of the incoming wave of the corresponding
mode.

Approximately, the time dependence of $M$ and $J$ can be
determined by
\begin{eqnarray}
&&-\frac{d}{dt}\left(\begin{array}{c}M\\J\end{array}\right)\nonumber
\\
 &=&\frac{1}{2\pi}\sum_{s,l,m} g_s\int_0^\infty d\omega
   \frac{{}_s\Gamma_{l,m}(r_h,a;\omega)}{e^{(\omega-m\Omega)/T}\mp 1}
   \left(\begin{array}{c}\omega\\m\end{array}\right),
   \label{eq:time_dependence}
\end{eqnarray}
where $g_s$ is the number of `massless' degrees of freedom at
temperature $T$, namely the number of degrees of freedom whose
masses are smaller than $T$, with spin $s$.  Therefore, once we
obtain the greybody factors, we completely determine the Hawking
radiation and the subsequent evolution of the black hole up to the
Planck phase, at which the semi-classical description by the
Hawking radiation breaks down and a few quanta radiated is not
predictable.

\subsection{Greybody factors for Randall-Sundrum black hole}
We find analytic expression of the greybody factors for $n=1$
Randall-Sundrum black hole within the low frequency expansion.
Here we outline our procedure: First we obtain the ``near
horizon'' and ``far field'' solutions in the corresponding limits;
Then we match these two solutions at the ``overlapping region'' in
which both limits are consistently satisfied; Finally we impose
the ``purely ingoing'' boundary condition at the near horizon side
and then read the coefficients of
``outgoing'' and ``ingoing'' modes 
at the far field side. The ratio of these two coefficients can be
translated into the absorption probability of the mode, which is
nothing but the greybody factor itself.

First for convenience, we define dimensionless quantities
\begin{eqnarray}
\xi      = \frac{r-\rH}{\rH}, \hspace{0.3cm} \omtil   = \rH\omega,
\hspace{0.3cm} \Qtil    = \frac{\omega-m \Omega}{2\pi T}.
\end{eqnarray}

Matching the NH and FF solutions in the overlapping region
$1+|\Qtil|\ll \xi \ll 1/\omtil$, we obtain
\begin{eqnarray}
R_{\infty}=Y_{\rm
in}e^{-i\omtil\xi}\left(\frac{\xi}{2}\right)^{-1}
          +Y_{\rm out}e^{i\omtil\xi}\left(\frac{\xi}{2}\right)^{-2s-1},
\end{eqnarray}
where
\begin{eqnarray}
&{}&Y_{\rm in}\nonumber \\
&=&\frac{\Gamma(2l+1)\Gamma(2l+2)}{\Gamma(l-s+1)\Gamma(l+s+1)}
   \frac{\Gamma(1-s-i\Qtil)}{\Gamma(l+1-i\Qtil)}
   (-4i\omtil)^{-l+s-1}\nonumber \\
&&\mbox{}+
   \frac{\Gamma(-2l)\Gamma(-2l-1)}{\Gamma(-l-s)\Gamma(-l+s)}
   \frac{\Gamma(1-s-i\Qtil)}{\Gamma(-l-i\Qtil)}
   (-4i\omtil)^{l+s},\nonumber \\
&{}& Y_{\rm out} \nonumber
\\
&=&\frac{\Gamma(2l+1)\Gamma(2l+2)}{[\Gamma(l-s+1)]^2}
   \frac{\Gamma(1-s-i\Qtil)}{\Gamma(l+1-i\Qtil)}
   (4i\omtil)^{-l-s-1}\nonumber \\
&&\mbox{}+
   \frac{\Gamma(-2l)\Gamma(-2l-1)}{[\Gamma(-l-s)]^2}
   \frac{\Gamma(1-s-i\Qtil)}{\Gamma(-l-i\Qtil)}
   (4i\omtil)^{l-s}.
\label{eq:coeff_xi}
\end{eqnarray}

Finally, the greybody factor $\Gamma$ (=the absorption
probability) could be written as follows.
\begin{eqnarray}
\Gamma =1-\left|\frac{Y_{\rm out}
  Z_{\rm out}}{Y_{\rm in}Z_{\rm in}}\right|
=1-\left|\frac{1-C}{1+C}\right|^2, \label{eq:Page_trick}
\end{eqnarray}
where
\begin{eqnarray}
C=\frac{(4i\omtil)^{2l+1}}{4}\left(\frac{(l+s)!(l-s)!}{(2l)!(2l+1)!}\right)^2
   \left(-i\Qtil-l\right)_{2l+1},
\end{eqnarray}
with $(\alpha)_n=\prod_{n'=1}^n(\alpha+n'-1)$ being the
Pochhammer's symbol.

For concreteness, we write down
the explicit expansion of eq.~(\ref{eq:Page_trick}) up to
$O(\omtil^6)$ terms
\begin{eqnarray}
{}_0\Gamma_{0,0}&=&4\omtil^2-8\omtil^4+O(\omtil^6),\nonumber\\
{}_0\Gamma_{1,m}&=&\frac{4\Qtil\omtil^3}{9}\left(1+\Qtil^2\right)
   +O(\omtil^6),\nonumber\\
{}_0\Gamma_{2,m}&=&\frac{16\Qtil\omtil^5}{2025}\left(
               1+\frac{5\Qtil^2}{4}+\frac{\Qtil^4}{4}\right)
               +O(\omtil^{10}),\nonumber
\end{eqnarray}
\begin{eqnarray}
{}_{\frac{1}{2}}\Gamma_{\frac{1}{2},m}
   &=&\omtil^2\left(1+4\Qtil^2\right)
   -\frac{\omtil^4}{2}\left(1+4\Qtil^2\right)^2
   +O(\omtil^6),\nonumber\\
{}_{\frac{1}{2}}\Gamma_{\frac{3}{2},m}
   &=&\frac{\omtil^4}{36}\left(
      1+\frac{40\Qtil^2}{9}+\frac{16\Qtil^4}{9}\right)+O(\omtil^8),
      \nonumber
\end{eqnarray}
\begin{eqnarray}
{}_1\Gamma_{1,m}
   &=&\frac{16\Qtil\omtil^3}{9}\left(1+\Qtil^2\right)+O(\omtil^6),\nonumber\\
{}_1\Gamma_{2,m}
   &=&\frac{4\Qtil\omtil^5}{225}\left(
               1+\frac{5\Qtil^2}{4}+\frac{\Qtil^4}{4}\right)
               +O(\omtil^{10}). \label{eq:Gamma_explicit}
\end{eqnarray}
Note that subleading terms in $\omtil$ are already neglected when
we obtain eq.~(\ref{eq:Page_trick}) and the terms from these
contributions are not written nor included in
eqs.~(\ref{eq:Page_trick}) and (\ref{eq:Gamma_explicit}). We also
note that the so-called s-wave dominance is maximally violated for
spinor and vector fields since there are no $l=0$ modes for them.

\bigskip

\section{Summary}
We have studied theoretical aspects of the rotating black hole
production and evaporation.

For production, we present an estimation of the geometrical cross
section 
up to unknown mass and angular momentum loss in the balding phase.
Our result of the maximum impact parameter $\bmax$ is in good
agreement with
the numerical result by 
Yoshino and Nambu when the number of extra dimensions is $n\geq 1$
(i.e.\ within 6.5\% when $n=1$ and 1.5\% when $n\geq 2$). Relying
on this agreement, we obtain the (differential) cross section for
a given mass and an angular momentum, which increases linearly
with the angular momentum up to the cut-off value
$J_\mathrm{max}=\bmax M/2$. This result shows that black holes
tend to be produced with large angular momenta. We also studied
the possibility of the black ring formation and find that it would
possibly form when there are many extra dimensions. For
evaporation, we first derive the master equation for brane fields
for general spin and for an arbitrary number of extra dimensions.
We show that the equations are separable into radial and angular
parts as the four-dimensional Teukolsky equations. From these
equations, we obtain the greybody factors for brane fields with
general spin for the five-dimensions ($n=1$) Kerr black hole
within the low-frequency expansion. We address several
phenomenological implications of our results. The form factor of
black hole production cross section is larger in the higher
dimensional spacetime. The more precise determination of the
radiation power is now available. We have shown that the black
holes are produced with large angular momenta and that the
resultant radiations will have strong angular dependence for
$s=1/2$ and $s=1$ modes which points perpendicular to the beam
axis while very small angular dependence is expected for scalar
mode. More quantitative estimation will need the greybody factors
for arbitrary frequency.
\vspace{-0.5cm}
\begin{acknowledgements}
We would like to thanks to G. 't Hooft for nice comments on our
work at DESY workshop and also to  Robert Emparan, Yiota Kanti, V.
Cardoso and M. Cavaglia for valuable discussions and comments. SC
would like to thanks to Don Page , Valeri Frolov, Maurice van
Putten and Eun-Joo Ahn for nice discussions at ICGA-VI. Y. Okada
and Danny Marfatia gave constructive comments to SC at ICFP.
\end{acknowledgements}
\vspace{-0.5cm}


\bibliography{paper}
\bibliographystyle{utphys}

\end{document}